\documentclass[11pt,a4paper]{amsart}

\usepackage{hyperref}
\usepackage{braket}
\usepackage{amssymb}
\usepackage{amsmath}
\usepackage{amsthm}
\usepackage{amscd}
\usepackage{mathptmx}
\usepackage{epigraph}

\newtheorem{theorem}{Theorem}[section]

\newtheorem{proposition}[theorem]{Proposition}

\theoremstyle{definition}

\theoremstyle{remark}

\begin{document}

\title{On some information-geometric aspects of Hawking radiation}

\author{Xiao-Kan Guo}
\address{
University of  Chinese Academy of Sciences, Beijing 100049, China}
\email{kankuohsiao@whu.edu.cn}
\thanks{The published version of this paper \cite{Guo15} contains lots of typos and imprecise statements. In this version we will add footnotes "{\it Notes added to the pulished version}..." to correct the wrong/ambiguous statements.}


\date{\today}


\begin{abstract}
This paper illustrates the resemblance between the information-geometric structures of probability spaces and that of the discrete spectrum for Hawking radiation. The information geometry gives rise to a reconstruction of the standard formalism of quantum mechanics, while the discrete spectrum of Hawking radiation contributes to the semiclassical unitary evolution of Hawking radiation. If more realistic models of Hawking radiation are chosen, the information-geometric structures of the probability space for Hawking radiation can be constructed from some physical considerations. The constructed quantum formalism is consistent with both the unitary evolution of Hawking radiation in the semiclassical picture and the topology change of fuzzy horizons. These aspects of Hawking radiation can be connected to some general convictions of quantum gravity. A comparison with the fuzzball proposal shows the limiation and effectiveness of this construction. We conclude that these information-geometric aspects show some possible ways bridging the gap between semiclassical models and quantum gravity. 
\end{abstract}

\maketitle

\epigraph{{\it Indeed, really new physical ideas usually require a new
mathematical formalism with which to describe them....quantum field theory is not everything. Nature's richness is not bounded by quantum field theory.}}%
{Xiao-Gang Wen\\ {\it Quantum Field Theory of Many-body Systems}\\(OUP, 2004, p.ix)}
\section{Introduction}
Quantum field theory in curved spacetime predicts many intriguing physical results, such as the Hawking radiation and the Unruh effect \cite{Bire82}. However, the ideal mathematical abstractions in the field-theoretic analysis prevent it from corresponding directly to the phenomena in the realistic physical world. For instance, boxes with ideally perfect mirrors as box walls experiencing the Unruh effect will cause the alleged self-accelerating box paradox, which will be greatly suppressed if a more realistic model of mirror is chosen \cite{PhysRevD.66.104004}. In the case of Hawking radiation, the information loss paradox is a similar problem. Ideally field-theoretic evaluations of the Bogoliubov coefficients show that the radiation spectrum is purely thermal and hence the nonthermal information is lost if the black hole evaporates completely \cite{PhysRevD.14.2460}. But if we consider the Hawking radiation as a tunneling process and take into account the back reaction of the radiation \cite{PhysRevLett.85.5042}, we can not only easily obtain the related quantities in black hole physics, but also have a more realistic picture of Hawking radiation with spectrum slightly differing from the original field-theoretic calculation.\footnote{{\it Notes added to the published version}: There is a notable exception \cite{PS09} that the Hawking radiation from linear dilaton black holes in Einstein-Maxwell-Dilaton theory calculated by the tunneling method remains thermal! One way out is to referring to quantum gravitational corrections (or other corrections from realistic astrophysical sources) to the tunneling formalism \cite{SHP11} so that the hidden-messenger arguments still apply. However, there is still a exception that the pair creation of Kaluza-Klein particles cannot be consistently described by a tunneling process \cite{FV05}, which is ignored/neglected by almost all the black-hole tunneling papers. In fact, most of the derivations of Hawking radiation do not refer to the conception of {\it particles}, and on the other hand backreaction can be implemented in those models in one way or another. Therefore, now I admit that the tunneling picture is not a universal one, and hence the assertions of this paper are mostly speculations! }
In particular, recent works have argued that this slight difference allows the \textit{hidden messengers} \cite{Zhang200998}, i.e. the correlations between succesive radiations, to recover  nonthermal information from black holes.\footnote{{\it Notes added to the published version}: The hidden-messenger proposed by Zhang-Cai-You-Zhan is a very general argument since it only depends on the tunneling rate of the form $e^{\Delta S}$ which is shown to hold for all tunnelings \cite{BP11}. However, it appears that such a general argument based on $e^{\Delta S}$ was first used by the Beijing Normal University group in 2008 \cite{ZDL08} before the Zhang-Cai-You-Zhan paper in 2009 \cite{Zhang200998}! }
 Further, it is shown that the average effects of the dynamical processes of Hawking radiation lead to an effective formulation of the black hole quasi-normal modes that takes a discrete and countable form such that a radiating black hole looks like an excited atom \cite{epjc/s10052-013-2665-6}, which results in a unitary evolution model for Hawking radiation described by a time-dependent Schr\"odinger equation \cite{1304.1899}.

All these advances listed above are carried out in a picture that has been modified constantly to be more realistic instead of a picture employing certain delicate, of course more undetectable, quantum gravity model. One might argue that these reality-oriented modifications might make the picture become more classical, whence the assertion that the information is conserved during the Hawking radiation is unconvincing at the semiclassical level. However, in spite of this drawback, the semiclassical unitary evolution of Hawking radiation at this level should reveal some features of the underlying full quantum theory. One foremost feature in \cite{epjc/s10052-013-2665-6} is the discreteness of the horizon area as the function of the overtone number of quasi-normal modes, which is consistent with various quantum gravity models where the spacetime is fundamentally descrete (see, e.g. \cite{lrr-1998-13}).

On the other hand, Hawking radiation is theoretically predicted and can be detected \textit{in principle}. While \textit{in practice}, this prediction is still far from being confirmed. Till now the most famous detecting proposal is the Unruh-DeWitt detector \cite{Bire82}. The Unruh-Dewitt detector is a theoretically designed point detector interacting locally with a quantum field, which is originally used to prove the existence of Unruh effect for detectors moving in Minkowski spacetime. In fact, a static detector can be utilized in black hole spacetime to detect the thermal spactrum of Hawking radiation \cite{PhysRevD.13.2188}. By adding some realistic considerations to the detector, such as the spatial profile of the detector and the finite time effect, we can get results consistent with locality by some peculier regularizations \cite{0264-9381-23-22-015}. More realistic models of the spatially smeared detectors can be constructed by putting detecting atoms in standard QED interactions \cite{PhysRevD.87.064038}. In fact, Hawking radiation can be attributed to the Lamb shifts of the collapsing atoms provided that only the high-energy modes can be really detected \cite{0264-9381-14-1A-024}. With these realistic models in mind, we can try to identify the tunneling rate of  Hawking radiation with the spectral response (or the transition rate) of the Unruh-DeWitt detecter by this atom-atom model correspondence.\footnote{{\it Notes added to the published version}: Although this section aims to argue the realistic meaning of the approach, a truely realistic correspondence should match in the magnitude of physical quantities in additon to the formal analogies. This critictism is due to Prof. Xin-Zhou Li.}

For all these reasons we observe that, if the identification is appropriate, the probability space of the tunneling rates (or the responses) has alike geometric structures of those analyzed in \cite{Reginatto}, wherein a probability space (or probabilty simplex) $\{P^i;i=1,\dotsc,n,\sum_i P^i=1\}$ with an information metric
\begin{equation}\label{1}
ds^2=G_{ij}dP^i dP^j=\frac{\alpha}{2P^i}\delta_{ij}dP^i dP^j,
\end{equation}
can be set in motion by introducing $S^i$, the canonical conjugations of $P^i$, to form a symplectic structure. Subsequent attempts to extend the n-dimensional information metric \eqref{1} to the 2n-dimensional phase space $\{P^i,S^i\}$ require a K\"ahler structure on the phase space. After changing $\{P^i,S^i\}$ to the complex coordinates $\{\psi^i,\bar{\psi}^i\}$ via the Madelung transformation, the coordinate variables take the forms of wave functions in standard quantum mechanics. The underlying Hilbert space can thus be constructed from the symplectic form on this K\"ahler manifold. The interpretation of the quantity $S^i$ needs our further attention.\footnote{{\it Notes added to the published version}: In former versions I insisted that the interpretation of $S^i$ is controversial. But now I admit that the arguments in later sections are not very strong. } 

In this paper, we first, by considering special cases in which the dimension $n$ is odd, find that part of $S^i$ might carry the information of the topological structures of the pertinent K\"ahler manifold. Then in the context of evaporating black holes, we find that the standard quantum mechanical structures constructed from the probability space of the tunneling rates of Hawking quanta (or the responses of detectors) are consistent with the aforementioned  untarity of Hawking radiation in the semiclassical models, since with $S^i$ appearing on the exponent we can only detect the probability space $\{P^i\}$. Hence the effectively unitary quantum evolution can be envisioned as an emergent result from the coarse graining of a more fundamental fine-grained theory. We compare this emergent result with  the S-matrix for Hawking radiation proposed in \cite{'tHooft1990138} for Hawking radiation. Furthermore, these results are compared with the predictions of the fuzzball proposal, which exhibits limitations of this construction. Altogether, we conclude that these information-geometric aspects of Hawking radiation implies some possible connections between the semiclassical results and the deeper quantum gravity predictions.

This paper is organized as follows: In section \ref{sec2}, after briefly revisiting the information-geometric reconstruction of quantum mechanics, we show that part of $S^i$ are related to the nonvanishing first Chern class of the resultant K\"ahler manifold. In section \ref{sec3}, the reconstruction procedure is applied to evaporating black holes in the semiclassical picture and the physical comprehensions of this procedure are pointed out. In section \ref{sec4}, the reconstruction is compared with some general implications from quantum gravity, such as the S-matrix of Hawking radiation and fuzzballs. Section \ref{sec5} concludes.

\section{Information-Geometric Reconstruction of Quantum Mechanical Formalism}\label{sec2}
Information geometry is the study of probability space with a {\it metric} by using the tools from differential geometry.
In this section, we first outline the geometrically inspired reconstruction of quantum mechanics from the metric \eqref{1} \cite{Reginatto}.\footnote{There exist other reconstructions based on physical postulates which also start from \eqref{1} \cite{Mehrafarin,PhysRevA.78.052120}.} Then we give some remarks concerning the interpretation of the reconstructed formalism.

\subsection{Reconstruction}
Let us begin with a probability space $\{P^i;i=1,\dotsc,n,\sum_i P^i=1\}$ with the information metric \eqref{1}. The reconstruction consists of four steps:
\begin{enumerate}
\item First consider the case in which the probabilities change with time. Suppose that the evolutions of the probabilities can be generated by an action principle and introduce the $\{S^i\}$ conjugate to $\{P^i\}$. The Poisson brackets are defined in the usual sense with the symplectic matrix
\[
\Omega=\begin{pmatrix}0&{\bf 1}_{n\times n}\\-{\bf 1}_{n\times n}&0\end{pmatrix}.
\] 
Then the phase space $\{P^i,S^i\}$ becomes a symplectic manifold with the canonical symplectic structure $\omega=\sum_{i=1}^n dS^i \wedge dP^i$. And there exists a hamiltonian $H$ such that the  equations of motion are $\dot{P}^i=\{P^i,H\},\dot{S}^i=\{S^i,H\}$.
\item Since the metric $G_{ij}$ in \eqref{1} is defined on the n-dimensional configuration space $\{P^i\}$ only, whereas the evolving phase space $\{P^i,S^i\}$ is 2n-dimensional, we have to extend $G_{ij}$ to the 2n-dimensional $g_{ab}$. Consistency of the sympletctic structure and the metric structure requires the following 
\[
\Omega_{ab}=g_{ac}J^c_{\phantom{c} b},\quad J^a_{\phantom{a} c}g_{ab}J^b_{\phantom{b} d}=g_{cd},\quad J^a_{\phantom{a} b}J^b_{\phantom{b} c}=-\delta^a_{\phantom{a} c.}
\]
That is, the manifold must have a K\"ahler structure with complex structure $J_{ab}$. As a result, the extended metric $g_{ab}$ and the complex structure $J_{ab}$ become repectively
\[
[g_{ab}]=\begin{pmatrix}G&A^T\\A&(1+A^2)G^{-1}\end{pmatrix},\quad [J^a_{\phantom{a} b}]=\begin{pmatrix}A&(1+A^2)G^{-1}\\-G&-GAG^{-1}\end{pmatrix},
\]
where $G=[G_{ij}]$ and $A^T=GAG^{-1}$. It can be further shown that the information metric \eqref{1} is the only metric invariant under the congruent embedding by a Markov mapping  \cite{Cam86}, which constrians that $A$ must be a constant matrix $A=a\textbf{1}_{n\times n}$. 
\item Now we can perform a (modified) Madelung transformation to the complex coordinates,
\begin{align}
\psi^k&=\sqrt{P^k}\exp[i(\frac{\Lambda}{\alpha} S^k-\gamma \ln\sqrt{P^k})],\nonumber\\
\bar{\psi}^k&=\sqrt{P^k}\exp[-i(\frac{\Lambda}{\alpha} S^k-\gamma \ln\sqrt{P^k})],\label{8}
\end{align}
where $\Lambda=\frac{1}{1+A^2}$ , $\gamma=\frac{-A}{1+A^2}$ and $k=1... n$. Consequently,
\begin{align*}
[\Omega_{ab}]&=\begin{pmatrix}0&i\alpha\Lambda^{-1}\textbf{1}\\-i\alpha\Lambda^{-1}\textbf{1}&0\end{pmatrix},\\
[g_{ab}]&=\begin{pmatrix}0&\alpha\Lambda^{-1}\textbf{1}\\\alpha\Lambda^{-1}\textbf{1}&0\end{pmatrix},\quad [J^a_{\phantom{a}b}]=\begin{pmatrix}-i\textbf{1}&0\\0&i\textbf{1}\end{pmatrix}.
\end{align*}
Let $\frac{\alpha}{\Lambda}=\hbar$, then \eqref{8} look similar to the wave functions in standard quantum mechanics. If $A=0$, then it is reduced to the flat-space case in which with the second term on the exponent vanishes \eqref{8} take the same form of the wave functions in standard quantum mechanics. \item The inner product defining a Hilbert space can then be contructed also in this flat-space case by Kibble's geometrized inner product \cite{Kib79}
\begin{equation}\label{ip}
\braket{\psi|\varphi}=\frac{1}{2}\sum_i (\psi^i ,\bar{\psi}^i)(g+i\Omega)\begin{pmatrix}\varphi^i\\\bar{\varphi}^i\end{pmatrix}=\sum_i \bar{\psi}^i\varphi^i.
\end{equation}
\end{enumerate}
\subsection{Remarks}
Now an immediate question is that, since the above reconstruction does not correspond in an obvious way to the geometrization in \cite{Kib79}, can we directly use the inner product \eqref{ip}?\footnote{{\it Notes added to the published version}: This problem is studied later in my master thesis \cite{GuoTh}.} To answer this question, let us recall the ideas of Kibble's geometrization of quantum mechanics \cite{Kib79}: Let $H$ be a complex Hilbert space and $K$ a dense linear subspace of $H$. Denote by $K_0$ the set of all the nonzero vectors in $K$, then the projective space $\Sigma=K_0/U(1)$ is the {\it quantum phase space}. Suppose $\Sigma$ is a paracompact manifold based on a Fr\'echet-Shwartz space $V$ and is endowed with a weak symplectic strucutre $\omega$. (By weak we mean that if for all $Y\in T_u\Sigma$ where $T_u\Sigma$ is the tangent space at $u$ on $\Sigma$ one has $\omega_\mu(X,Y)=0$, then $X=0$.) $\omega$ defines a map from the tangent bundle to the cotangent bundle
\[\mu:T\Sigma\rightarrow T^*\Sigma;~X\mapsto X^\flat=i_X\omega\]
where $i_X\omega(Y)=\omega(X,Y)$ is the interior product. Let $\pi:K_0\rightarrow\Sigma$ be the projection to the quantum states represented by the vectors in $K_0$ satisfying $\pi{\bf u}=u$ in a local chart. For the normalized vectors ${\bf v}$ in $K_0$, the corresponding quantum states can always be represented by vectors ${\bf u}$ on a hypersurface $K_v=\{{\bf u}\in K_0;\braket{\bf v,u}=1\}$ in the neighborhood of $\pi{\bf v}$. Therefore, there exists a homeomorphism $\phi$ from this neighborhood to an open set in $K_{\bf v}$ such that
\[\omega_{\pi{\bf v}}=2\Im\braket{\phi_*X,\phi_*Y},\quad\forall X,Y\in T_{\pi{\bf v}}\Sigma
\]
where $\phi_*$ is the Jacobi map on the tangent space. Take a vector field $k$ on $\Sigma$, then $k$ generates the Hamiltonian flow $\tau_t=\exp(tK)$. The interal lines of theis flow are the history states which when pulled back from $\Sigma$ to $K$ give the dynamical Schr\"odinger equation. The inner product \eqref{ip} is then defined on the complexified tangent space $T_v^{\mathbb{C}}$.

We can now construct the geometrization of quantum mechanics in the statistical manifold.
\begin{proposition}
The quantum phsae space $\Sigma=(P,S)$ reconsctructed above is the cotangent bundle of the statistical manifold defined by the metric \eqref{1}. The corresponding tangent space realizes Kibble's geometrization of quantum mechanics in information geometry.
\end{proposition}
\begin{proof}(Sketch.) First, the information metric \eqref{1} is defined in the tangent space of the positive cone $\mathbb{R}^+$ of the probability simlex $S_{n-1}$. from the reconstruction one knows that $(P,S)$ is already a phase space with simplectic structure. The configuration space is the simplex $S_{n-1}$, and the tangent space is $(P_i,\dot{P}_i)$, whence the $(P,S)$ is the cotangent space of $S_{n-1}$.

Second, let us consider the realization of  Kibble's geometrization of quantum mechanics in information geometry. Take the probability simplex $S_{n-1}$ as the $K_0$, then the quantum phase space $\Sigma$ is the reduced simplex $S_{m-1},2\leqslant m\leqslant n$ (where by reduced we mean the indistinguishability between some of the $P_i$). Then from \cite{Cam86} one knows that there is a congruent markov map $\phi$ from $S_{m-1}$ to $S_{n-1}$ such that the metric \eqref{1} is invariant. Similarly, the Markov map $\phi$ on $\Sigma$ induces the Jacobi map $\phi_*$  and defines the geometric inner  product \eqref{ip} giving rise to the weak symplectic structure $\omega$ on $\Sigma$. The weak symplectic structure $\omega$ then defines the map $\mu$.
\end{proof}
To define complex structure on the cotangent space $T^*$, one must complexify it to $T^*\otimes\mathbb{C}$. This point is not stated in the above reconstruction steps, but it is achievable due to the following information-geometric version of the Dombrowski theorem of Riemannian geomtry.
\begin{theorem}[Molitor \cite{Mol13}]
For an exponential family of probabilities $E=\{p(x;\theta)\}$ on a measure space $(\Omega,dx)$, where 
\[p(x;\theta)=\exp\Bigl\{C(x)+\sum_{i=1}^{n}\theta_iF_i(x)-\psi(x)\Bigr\},\quad C,F:\text{measurable functions},\theta_i\in\mathbb{R},
\]
its tangent bundle is a K\"ahler manifold.
\end{theorem}
Since the integral lines on a statistical manifold represents an exponential family of probabilities \cite{Cam85}, we see that the use of inner product is reasonable.\footnote{{\it Notes added to the published version}: In \cite{Mol13} the obervables, spectra of the obervables and the probabilty distributions are constructed from the K\"ahler function.}

Next, let us discuss the meaning of  $S^i$. In the reconstruction steps the $S^i$ are coordinates of the cotangent bundle conjugate to the $P_i$. Althogh the symplectic form $\omega$ defined the map $\mu$ from the tangent vector fields to the cotangent vector fields, the $\mu$ is not necessarily a bijection since $\omega$ is weak and it actually depends on the topology of the base space $V$. Moreover, the modulus square of the formal wave functions \eqref{8} gives us  the probability $P_i$ without the factor $e^{iS_i}$. Since the global U(1) phase has already been quotiented out in the definition of quantum phase space $\Sigma$, one might think this factor is similar to  the plane wave wave function $e^{ipx}$ and $S=px$. However, this analogy fails for other quantum states, e.g. the Gaussian state $\sim e^{ikx-x^2/(2d)^2}$. Hence $S$ contains more information than we expect.

Let us discuss the topological meaning of some part of $S^i$. On the one hand, the allowed tranformation group preserving the probabilty normalization and the information metric is $O(2n)$ and hence the vectors in the probability space can be connected by curves lie on the unit sphere $S^n$ \cite{Reginatto}. On the other hand, the observables, say the expectation values, should be invariant under the $U(1)$ phase shift. If the dimension of the unit sphere $n=2k+1$ is odd,\footnote{Note that when the dimension of the probabilty space is even it cannot be reduced to a $\mathbb{C}P^n$, thereby the above  interpretation is not generically correct. Nevertheless, the even cases can be assumed to have trivial topology. We speculate that the dimension of the probability space might be constrained  by its topology. Disregarding this speculation, one can still give the following arguments. The invalidity of the use of the second Chern class for the U(1) reduction of an even dimensional space has already been pointed out in \cite{Yoneya}. The monopoles, as will be shown to be the case of current interst in the next section, are only possible when the space dimension is odd, whereas when the dimension is even, one has instanton solutions, which deserve further investigations especially in the case of Hawking radiation.} 
then  the probability space is just the complex projective space $\mathbb{C}P^k\simeq S^{2k+1}/U(1)$. It is well-known that $\mathbb{C}P^n$ is a K\"ahler manifold with the standard invariant Hermitian metric \cite{Hou}\footnote{{\it Notes added to the published version}: In fact, using homogeneous coordinates one can write this metric as the Fubini-Study metric which is the K\"ahler metric of $\mathbb{C}P^n$.}
\begin{equation}\label{13}
ds^2=\sum_{k=0}^n dz^k d\bar{z}^k-(\sum_{i=0}^n z^i d\bar{z}^i)(\sum_{j=0}^n \bar{z}^j dz^j),
\end{equation}
where $z^k,\bar{z}^k$ satisfying $\sum_k z^k\bar{z}^k=1$ are coordinates on $S^{2n+1}$. The second term in \eqref{13} is similar to the correction term added to the information metric \eqref{1} in \cite{Mehrafarin} identifying the disturbance of different preparations of the systems to the measurement results. Note that a given probability distribution can be interpreted as an explicit way of preparing the quantum state, and different environment of the preparation can be represented as different bases of the Hilbert space. Different probability distributions, or histories, have correlation between them and decohere very fast \cite{Zeh73}. Therefore the inereference is possible no matter how small they are. The corrected metric also leads to the complex coordinates in \cite{Mehrafarin},  \[\psi^i=\sqrt{P^i}\exp [i\phi^i]\] where $\phi^i$ is just the degrees of disturbance identified by the correction term.

From another point of view,  $\mathbb{C}P^n$ is also a K\"ahler-Einstein manifold with the Ricci form $\rho$ proportional to the K\"ahler form $\omega$ \cite{Hou}, \[\rho=(2n+2)\omega.\] On a K\"ahler manifold the Ricci form $\rho$ is identical to the first Chern class $c_1$. In the case of $\mathbb{C}P^n$, the K\"ahler form and hence the first Chern class does not vanish. That is to say, the holonomy group cannot be reduced to $SU(n)$ due to the topological obstructions. We see that $S^i$ can be really "topological".\footnote{{\it Notes added to the published version}: Two paragraphs in the followng are deleted since most of the discussions are meaningless.}

Note that in \cite{Reginatto} the phase factors $S^i$ are related to the classical groups of motion as the momentum densities, so that the formal wave functions can be interpreted as the wave functions in B\"ohmian mechanics. This interpretation is of course desirable for the classical limit, however, it is not the whole story in quantum phenomena. For instance, after a measurement of a quantum state, what one obtains is the probability distribution $P^i$ instead of $S^i$, but does one lose only the information of particle momenta? The answer is obviously no. On the other hand, the quantum potentials of B\"ohmian mechanics only appears in the equations of motion, and whether the $S$ are related to the Hamiltonian principal fuction in the Hamilton-Jacobi euqations is still unknown.

We can understand this topological interpretation  also from the universal coefficient theorem and the resultant principle of topological covariance \cite{PhysRevD.90.045018}. Here the change in topology is described by the change in the coefficient groups in the cohomology $H^2(G,U(1))$ of the probalility space, which are probed by particular experimental setups. The phase factors $\exp(iS^i)$ in \eqref{8} characterize the group extension $G/U(1)$ of the related group $G$ by $U(1)$. By the universal coefficient theorem
\[
0\rightarrow\text{Ext}(H_1(G,R),U(1))\rightarrow H^2(G,U(1))\rightarrow \text{Hom}(H_1(G,R),U(1))\rightarrow0,
\]
there exist different elements in $H^2(G,U(1))$ mapped into the same element in Hom, and the extension Ext chrachterizes different choices of coefficeint structure $R$. Hence the measured probability distributions are identical for different states up to extension factors $\exp(iS^i)$. Since the group extension factors $\exp(iS^i)$ depends on the choice of the coefficient groups, we see that $S^i$ indeed carry the information about the topological structrues of the probability space.
\section{Quantum mechanics of Hawking radiation}\label{sec3}
Now we can apply the above results to the situation of evaporating black holes emitting Hawking radiation to construct the quantum mechanics of black holes.\footnote{{\it Notes added to the published version}: The reason to do this is that technically Hawking radiation is a quantum effect since the radiated particles are microscopic, and we can have a statistical mechanical understanding as the microscopic corrections (e.g. fluctuations) to the macroscopic behaviors \cite{Witten12}. Hence it is meaningful to do so.}

As mentioned above, the Hawking radiation spectrum is discrete in the effective tunneling model \cite{epjc/s10052-013-2665-6}, which is consistent with the discrete spacetime in existing quantum gravity models \cite{lrr-1998-13}, especially the discrete Hawking radiation spectrum leads to the discrete horizon area. Consider now the emission rate of Hawking radiation in the semiclassical tunneling picture
\begin{equation}\label{14}
\Gamma_{\text{out}}\sim\exp[\Delta S]=\exp[\Delta A/4]
\end{equation}
which now becomes discrete. Notice that \eqref{14} is a probability rate, the probability of emission per unit time, instead of the probability. However, we further notice that in the unspecified actual probability space the different discrete probability rates indicate that this probability space is discrete and is changing or evolving with time. This is exactly the type of the  probability space that we were discussing in section \ref{sec2} with the trivial normalization. The information metric \eqref{1} can be considered as a consequance of the statistical distance between different probabilistic outcomes \cite{PhysRevD.23.357} and hence exists for this probability space.\footnote{{\it Notes added to the published version}: The mathematical arguments for this existence can be found in \cite{Mol13}.}

Now that the probability space $\{P_{\text{out}}^i\}$ of Hawking radiation is evolving with time, we have to ask whether the laws of motion of this probability space is just what we have constructed in section \ref{sec2}. At the first sight, the evolution of the Hawking radiation probability should be attributed to certain underlying physics and simply enforcing additional quantities such as $S^i$ will not be useful at all.  However, we can turn to the observation side to see what the Unruh-DeWitt detectors tell us.

Before we proceed it is worth noticing that the transition rate $w$ of an Unruh-DeWitt detector  is defined in terms of the time derivative of the response function $F(\omega,\tau)$ as \cite{Bire82}
\begin{align}
w=&\lambda^2 \mu^2 \dot{F}(\omega,\tau),~\text{with}\label{15}\\
F(\omega,\tau)=&\int_{\tau_0}^{\tau}d\tau^{\prime}\int_{\tau_0}^{\tau}d\tau^{\prime\prime}e^{-i\omega(\tau^{\prime}-\tau^{\prime\prime})}\cdot\bra{0}\phi(\tau^{\prime})\phi(\tau^{\prime\prime})\ket{0}\nonumber
\end{align}
if the monopole interaction $H_I=\lambda\mu\phi$ is assumed. After some calculations we can see that the transition rate takes the form of thermal spectrum
\begin{equation}\label{17}
\dot{F}(\omega,0)=\frac{1}{2\pi}\frac{\omega}{e^{2\pi\omega}-1},
\end{equation}
which seems to be different from the tunneling rate \eqref{14}. In fact, the transition rate \eqref{17} has averaged over an ensemble of atoms in the detector, while the tunneling rate \eqref{14} is for a single Hawking particle. This can be reconciled by evaluating the spectrum corresponding to \eqref{14}, as is carried out in \cite{Banerjee2009243}. If the condition of detailed balance is reached, \eqref{14} and \eqref{17} give the same Boltzmann factor
\begin{equation}
\frac{\dot{F}(\omega,\infty)}{\dot{F}(-\omega,\infty)}=\frac{\Gamma_{out}}{\Gamma_{in}}=\exp[-\beta\omega]
\end{equation}
with temperature $1/\beta$. 

In the light of this observation we expect the evolution of $P_{\text{out}}^i$ can be inferred from that of $\dot{F}$. The evolutions of the detector's atoms are of course quantum mechanical and can be described by a recently presented approximate master equation for the density matrix of the detector \cite{PhysRevD.89.064024}
\begin{equation}\label{19}
\dot{\rho}_{mm}=\sum_{k\neq m}[\rho_{kk}w_{mk}-w_{km}\rho_{mm}]
\end{equation}
in terms of the matrix elements
with $w$ given by \eqref{15}. As we can see, the transition rate $w$ here behaves like a Hamiltonian for the evolution of the detector (atoms). Since $w(\dot{F},\cdot)$ is a function of $\dot{F}$'s or rather of $F$'s and other possible parameters, we could choose these unspecified parameters as quantities $G$'s conjugate to the $F$'s. Then by the fact  that once we have such a Hamiltonian function $w$ a Hamiltonian vector field can be constructed as
\begin{equation}
X_w=\sum_{\omega}(\frac{\partial w}{\partial F_{\omega}}\frac{\partial}{\partial G_{\omega}}-\frac{\partial w}{\partial F_{\omega}}\frac{\partial}{\partial G_{\omega}})
\end{equation}
from which the Poisson brackets come out immediately.
We thus have, as a direct consequence of the quantum mechanical equation of evolution \eqref{19}, the desired symplectic stucture for the evolving probabilty space $\{F_{\omega}\}$ for the clicks of the detector. Correspondingly, we have the same symplectic stucture for the probability space $\{P_{\text{out}}^i\}$ of Hawking radiation.\footnote{{\it Notes added to the published version}: Such an connection is a possibility but far from being correct. This connection comes from the similar physical descrition of i) the spatially smeared UD detector and ii) the Hawking radiation derived from the Lamb shifts of quantum harmonic oscillators. The explicit calculations can be found in e.g. \cite{ZY12}}

Starting from this symplectic sructure of $\{P_{\text{out}}^i\}$, we can follow the discussion in section \ref{sec2} step by step and obtain the complex coordinates of the pertinent K\"ahler manifold that are in the form of wave functions 
\begin{equation}\label{21}
\psi^i_{\text{out}}=\sqrt{P^i_{\text{out}}}\exp[\frac{i}{\hbar}S^i],~\bar{\psi}^i_{\text{out}}=\sqrt{P^i_{\text{out}}}\exp[\frac{-i}{\hbar}S^i],
\end{equation}
with $i=1,...,n$ the same as the probability space.

The phase factors $S^i$, as is illustrated above, carry the information about the topological structures of the probability space. This still can be correct in the current case of black holes.\footnote{{\it Notes added to the published version}: One way to understand the BH information problem is to borrow ideas from quantum cosmology that the topology change is lost in the third quantization \cite{Hsu08,Fai14}.} Recall that Hawking radiation as topology changes of a fuzzy sphere $S^2_F$ was proposed in \cite{Silva2009318}.\footnote{{\it Notes added to the published version}: From the perspective of string theory, a black hole can be formed by collapsing a fuzzy sphere \cite{IKRS13}. In this case, the fuzzy sphere is realized as the $D$-2 branes consisting of $D$-0 branes, and Hawking radiation can be described as tachyon condensation.} 
In this approach the fuzzy sphere Hilbert space is $(2J+1)$-dimensional and the selection rule for the area transition restricts to the following $J\rightarrow J-1/2$ (where $J^{a}$ is the n-dimensional irreducible representation of $\mathfrak{su}(2)$ Lie algebra and the $J$'s are half integers). We therefore see that the wave functions \eqref{21} can correspond to the wave functions $M$ defined on the fuzzy sphere. Since the fuzzy sphere is a noncommutative manifold, one has the coproduct in its Hopf algebra
\begin{align*}
\Delta: S^2_F(j)&\rightarrow S^2_F(k)\otimes S^2_F(l);\\
M&\mapsto\Delta(M)=\sum_{\mu_1,\mu_2,m_1,m_2}C_{k,l;\mu_1,\mu_2}C_{k,l;m_1,m_2}M_{\mu_1+\mu_2,m_1+m_2}e^{\mu_1m_1}\otimes e^{\mu_2m_2}
\end{align*}
where the $C$'s are CG coefficients. This is the process of Hawking radiation as topology change or the spliting of $M$ into the main and the baby world with the degrees of freedom in the baby world inaccessible to observers in the main world. In the language of the above reconstruction, is a measurement of the quantum states to obtain the probabilties with the phase factors that characterize the topological structures being lost. As a consequance, observers in the main world can not detect the change in the Hilbert space dimension or the fine-grained topology. What can be detected is the coarse-grained probabilility space under the quantum mechanical evolution.

In order to show more details we first recall that the algebra $\text{Mat}_n$ of $n\times n$ matrices is the structure algebra of $S^2_F$, which can be generated, as is chosen above, by the SU(2) irreducible representation $J^a$. In terms of eigenvectors of $J^3$
\[
J^3\ket{m}=m\ket{m},~m=-j\ldots j,~j=\frac{n-1}{2},
\]
we can define coherent states 
\[
\ket{\alpha}=(\frac{1}{2}\sin\theta)^j\sum_{m=-j}^j\sqrt{\frac{(2j)!}{(j+m)!(j-m)!}}(\frac{\alpha}{R})^{-m}\ket{m}
\]
where $\alpha=R\tan(\theta/2)\exp[i\phi]$ and $(R,\theta,\phi)$ are spherical coordinates. Then the fuzzy sphere $S^2_F$ contains $n$ cells each of which supports a coherent state $\ket{\alpha}$. With these $n$ coherent states one can choose orthonormal basis for polynomial functions on $S^2_F$ in terms of coupled oscillators $\{a_0^{(\dagger)},a_1^{(\dagger)}\}$,
\begin{equation}\label{2255}
\sqrt{\begin{pmatrix}n\\k\end{pmatrix}}a_0^{n-k}a_1^k.
\end{equation}
Then with the help of the projectors $p=\ket{\psi_n}\bra{\psi_n}$ determining projective modules, with \[\ket{\psi_n}=N\begin{pmatrix}a_0^n\\\ldots\\\sqrt{\begin{pmatrix}n\\k\end{pmatrix}}a_0^{n-k}a_1^k\\\ldots\\a_1^n\end{pmatrix}\] being the basis vector that consists of \eqref{2255}, the first Chern class can be calculated as in \cite{grs},
\begin{align}
c_1&=\frac{-1}{2\pi i}\int \text{Tr}p(dp)(dp)\neq0\nonumber\\
&\xrightarrow{n\rightarrow\infty} k \in\mathbb{Z}.\nonumber
\end{align}
Hence we see that the $n$ cells on  $S^2_F$ do carry topological charges that are in analog to the N-bit strings in each cell in \cite{zizzi}. And that the measurements change the topological charges is analogous to the measurements taking out a cell to get N classical bits. The phase factors are cancelled in the projector $p$ and are undetectable in conventional analysis. However, the fuzzy line bundles or equivalently the projective mudules are classified by the characteristic classes or topological charges, which corroborates the topological interpretations in the last section.

\section{Implications from the Reconstruction}\label{sec4}
In this section, we discuss the implications for quantum gravity in the above reconstructed formalism.\footnote{{\it Notes added to the published version}: The comparison with the 't Hooft's deterministic theory is deleted, since it is not oringinal. In \cite{GuoTh}, more examples are discussed suchas how to calculate black hole entropy on a fuzzy sphere and its relation to anyon statistics on a fuzzy sphere.}

Firstly,   in the above reconstructed Hilbert space, we can try to construct the S-matrix for Hawking radiation. For a single tunneling  of Hawking radiation from a Schwarzshcild black hole, the gravitational back reaction contributes to a phase shift in the tunneling rate \cite{PhysRevLett.85.5042}
\[
\Gamma_{\text{out}}\rightarrow\exp[4\pi\omega^2]\Gamma_{\text{out}}.
\]
Then for states $\ket{\psi^i_{\text{out}}}$ in the reconstructed Hilbert space, the phase shift becomes
$\exp[2\pi\omega^2]$.
Consider an ingoing state $\ket{\psi^i_{\text{in}}(\omega)}$ that contributes to the Hawking radiation of energy $\omega$. The outcoming state under back reaction is 
\[
\ket{\psi^i_{\text{out}}(\omega)}\rightarrow\exp[2\pi\omega^2]\ket{\psi^i_{\text{out}}(\omega)}.
\]
The S-matrix is then
\[
\braket{\psi^i_{\text{out}}(\omega)|\psi^i_{\text{in}}(\omega)}=N\exp[2\pi\omega^2]
\]
for some normalization factor $N$. This is consistent with the S-matrix of black hole radiation presented in \cite{'tHooft1990138},
\begin{equation}
\braket{\psi^i_{\text{out}}(\omega)|\psi^i_{\text{in}}(\omega)}\sim\exp[i\int d\omega p_{\text{in}}(\omega)p_{\text{out}}(\omega)].
\end{equation}

Secondly, an intriguing comparison with the fuzzball proposal for black holes can be made. The horizon-size horizonless fuzzballs are microstates of certain black holes in string theory. If all known black holes admit this microscopic structure, one has to find ways to connect it to the classical notions such as horizons. In \cite{0810.4525} it is conjectured that fuzzballs can be formed from the collapsing shells via phase evolution
\begin{equation}\label{28}
\sum_k c_k\ket{E_k}\rightarrow\sum_k c_k e^{-iE_k t}\ket{E_k}
\end{equation}
which takes the initial special superpositon to the general superposition that includes much more fuzzball states. This dephasing effect can be well illustrated by the fuzzball complementarity \cite{Mathur2014566} where high energy infalling quanta with $E\gg T_{\text{BH}}$ collectively excite the fuzzballs and experience coarse-grained free fall, while quanta with $E\sim T_{\text{BH}}$ experience the fine-grained structures of fuzzballs and do not feel free fall. Hence, complementarity does not hold for Hawking quanta with $E\sim T_{\text{BH}}$ that encodes the black hole information and for a distant Unruh-DeWitt detector the unitary detection is not necessarily equivalent to knowing everything about the microscopic states. As a result, for high energy modes the complementarity holds and the Lamb shifts arguments also can be applied, which entails the reconstruction above. For low energy modes, although the cancelling of phases is possible it has nothing to do with the unitary evolution detected by a distant detector due to the invalidity of the complementarity. Indeed, in this case information is already carried out by Hawking quanta with $E\sim T_{\text{BH}}$ from fuzzballs but the observed unitary evolution does not reflect microscopic structures of fuzzballs, just as acted by an inverse map of \eqref{28}, and hence the effective atom-like structure endures.

\section{Conclusion}\label{sec5}
In this paper, we have discussed the information-geometric aspects of Hawking radiation. The procedure of reconstructing quantum mechanical formalism from information geometry is applied to the cases of evaporating black holes. We find that this reconstruction not only explains the successes of the semiclassical tunneling models but resembles other novel models of Hawking radiation. In addition, every step of this procedure can be based on adequate physical grounds. Connections to deeper quantum deterministic theory are pointed out, whereas comparison with the fuzzball proposal shows the limitations of this reconstruction. Part of the quantities conjugate to the probabilities in the above construction are shown to characterize the topological structures of the pertinent space.\footnote{{\it Notes added to the published version}: A subsequent paragraph in the published version is deleted, since it is meaningless.}

Notice that it has been argued that small corrections to the thermality of Hawking effect do entail unitary evolutions but fail to give correct statistical behavior of the entropy \cite{0264-9381-26-22-224001}. However, the analysis in this paper still holds at the semiclassical level. For more detailed microscopic description, the orthodox quantum mechanics might not be sufficient and an understanding of deeper structures is required.

\bibliographystyle{amsalpha}

\begin{thebibliography}{99}
\bibitem{Bire82} N. D. Birell and P. C. W. Davies, \emph{Quantum Fields in Curved Space} (Cambridge University Press, Cambridge, 1982).
\bibitem{PhysRevD.66.104004} D. Marof and R. D. Sorkin, Perfect mirrors and the self-accelerating box paradox, Phys. Rev. D {\bf66}, 104004 (2002).
\bibitem{PhysRevD.14.2460} S. W. Hawking, Breakdown of predictability in gravitational collapse, Phys. Rev. D {\bf14}, 2460 (1976).
\bibitem{PhysRevLett.85.5042} M. K. Parikh and F. Wilczek, Hawking radiation as tunneling, Phys. Rev. Lett. {\bf85}, 5042 (2000).
\bibitem{Zhang200998} B. Zhang, Q. y. Cai, L. You and M. S. Zhan, Hidden messenger revealed in Hawking radiation: A resolution to the paradox of black hole information loss, Phys. Lett. B {\bf675}, 98 (2009).
\bibitem{epjc/s10052-013-2665-6} C. Corda, Black hole quantum spectrum, Eur. Phys. J. C {\bf73}, 2665 (2013).
\bibitem{1304.1899} C. Corda, Time dependent Schrödinger equation for black hole evaporation: No information loss, Ann. Phys. {\bf353}, 71 (2015).
\bibitem{lrr-1998-13} R. Loll, Discrete approaches to quantum gravity in four dimensions, Living Rev. Relativity {\bf1}, 13 (1998).
\bibitem{PhysRevD.13.2188} J. B. Hartle and S. W. Hawking, Path-integral derivation of black-hole radiance, Phys. Rev. D {\bf13}, 2188 (1976).
\bibitem{0264-9381-23-22-015} J. Louko and A. Satz, How often does the Unruh-DeWitt detector click? Regularization by a spatial profile, Class. Quantum Grav. {\bf23}, 6321 (2006);\\ A. Satz, Then again, how often does the Unruh-DeWitt detector click if we switch it carefully?, Class. Quantum Grav. {\bf 24}, 1719 (2007).
\bibitem{PhysRevD.87.064038} E. Mart\' in-Mart\' inez, M. Montero and M. del Rey, Wavepacket detection with the Unruh-DeWitt model, Phys. Rev. D {\bf87}, 064038 (2013).
\bibitem{0264-9381-14-1A-024} D. J. Raine and D. W. Sciama, Hawking radiation and dissipative quantum systems, Class. Quantum Grav. {\bf14}, A325 (1997).
\bibitem{Reginatto} M. Reginatto and M. J. W. Hall, AIP Conf. Proc. {\bf1443}, 96 (2012);  AIP Conf. Proc. {\bf1553}, 246 (2013); M. Reginatto, arXiv:1312.0429.
\bibitem{'tHooft1990138} G. 't Hooft, The black hole interpretation of string theory, Nucl. Phys. B {\bf335}, 138 (1990).
\bibitem{Mehrafarin} M. Mehrafarin, Quantum mechanics from two postulates, Int. J. Theor. Phys. {\bf44}, 429 (2005).
\bibitem{PhysRevA.78.052120} P. Goyal, Information-geometric reconstruction of quantum theory, Phys. Rev. A {\bf78}, 052120 (2008).
\bibitem{Hou} B.-Yuan Hou and B.-Yu Hou, \emph{Differential Geometry for Physicists} (Science Press, Beijing, 2004).
\bibitem{Yoneya} T. Yoneya, Topology of Euclidean Yang–Mills fields: Instantons and monopoles, J. Math. Phys. {\bf18}, 1759 (1977).
\bibitem{PhysRevD.90.045018} A. Patrascu, Quantization, holography, and the universal coefficient theorem, Phys. Rev. D {\bf90}, 045018 (2014).
\bibitem{PhysRevD.23.357} W. K. Wootters, Statistical distance and Hilbert space, Phys. Rev. D {\bf23}, 357 (1981).
\bibitem{Banerjee2009243} R. Banerjee and B. R. Majhi, Hawking black body spectrum from tunneling mechanism, Phys. Lett. B {\bf675}, 243 (2009).
\bibitem{PhysRevD.89.064024} M. Fukuma, S. Sugishita and Y. Sakatani, Master equation for the Unruh-DeWitt detector and the universal relaxation time in de Sitter space, Phys. Rev. D {\bf89}, 064024 (2014).
\bibitem{Silva2009318} C. A. S. Silva, Fuzzy spaces topology change as possible solution to the black hole information
loss paradox, Phys. Lett. B {\bf677}, 318 (2009).
\bibitem{grs} P. Valtancoli, Projectors for the fuzzy sphere, Mod. Phys. Lett. A {\bf16}, 639 (2001);\\ H. Grosse, C. W. Rupp and A. Strohmaier, Fuzzy line bundles, the Chern character and topological charges over the fuzzy sphere, J. Geom. Phys. {\bf42}, 54 (2002).
\bibitem{zizzi} P. A. Zizzi, A minimal model for quantum gravity, Mod. Phys. Lett. A {\bf20}, 645 (2005).
\bibitem{0810.4525} S. D. Mathur, Fuzzballs and the information paradox: A summary and conjectures arXiv:0810.4525.
\bibitem{Mathur2014566} S. D. Mathur and D. Turton, The flaw in the firewall argument , Nucl. Phys. B {\bf884}, 566 (2014).
\bibitem{0264-9381-26-22-224001} S. D. Mathur, The information paradox: A pedagogical introduction, Class. Quantum Grav. {\bf26}, 224001 (2009).
\bibitem{Guo15} X.-K. Guo, On Some information-geometric aspects of Hawking radiation as tunneling, Int. J. Theor. Phys. {\bf54}, 3699 (2015).
\bibitem{PS09} H. Pasaoglu, I. Sakalli, Hawking radiation of linear dilaton black holesin various theories, Int. J. Theor. Phys. {\bf48}, 3517 (2009).
\bibitem{SHP11} I. Sakalli, M. Halilsoy, H. Pasaoglu, Entropy conservation of linear dilaton black holes in quantum corrected Hawking radiation, Int. J. Theor. Phys. {\bf50}, 3212 (2011).
\bibitem{FV05} T. Friedmann, H. Verlinde, Schwinger pair creation of Kaluza-Klein particles: Pair creation without tunneling, Phys. Rev. D {\bf71}, 064018 (2005).
\bibitem{BP11} S. L. Braunstein, M. K. Patra, Black hole evaporation rates without spacetime, Phys. Rev. Lett. {\bf107}, 071302 (2011).
\bibitem{ZDL08} Y.-P. Zhang, Q. Dai, W.-B. Liu, Unthermal Hawking radiation from a general stationary black hole, Commun. Theor. Phys. {\bf49}, 379 (2008).
\bibitem{Cam86} L. L. Campbell, An extended Cencov characterization of the information metric, Proc. Amer. Math. Soc. {\bf98}, 135 (1986).
\bibitem{Kib79} T. W. B. Kibble, Geometrization of quantum mechanics, Commun. Math. Phys. {\bf65}, 189 (1979).
\bibitem{GuoTh} X.-K. Guo, From information to black hole, Master Thesis, Shanghai Normal University, 2016.
\bibitem{Mol13} M. Molitor, Exponential families, K\"ahler geometry and quantum mechanics, J. Geom. Phys. {\bf70}, 54 (2013).
\bibitem{Cam85} L. L. Campbell, The relation between information theory and the differential geometry approach to statistics, Inf. Sci. {\bf35}, 199 (1985).
\bibitem{Zeh73} H. D. Zeh, Towards a quantum theory of measurement, Found. Phys. {\bf3}, 109 (1973).
\bibitem{Witten12} E. Witten, Quantum mechanics of black holes, Science {\bf337}, 538 (2012).
\bibitem{ZY12} W.-t. Zhou, H.-w. Yu, Spontaneous excitation of a static multilevel atom coupled with electromagnetic vacuum fluctuations in Schwarzschild spacetime, Class. Quantum Grav. {\bf29}, 085003 (2012).
\bibitem{Hsu08} S. D. H. Hsu, Spacetime topology change and black hole information, Phys. Lett. B {\bf644}, 67 (2008).
\bibitem{Fai14} M. Faizal, Absence of black holes information paradox in group field cosmology, Int. J. Geom. Methods Mod. Phys. {\bf11}, 1450010 (2014).
\bibitem{IKRS13} N. Iizuka, D. Kabat, S. Roy, D. Sarkar, Black hole formation in fuzzy sphere collapse, Phys. Rev. D {\bf88}, 044019 (2013).
\end{thebibliography}

\end{document}